\documentclass[11pt]{article}
\pdfoutput=1

\usepackage{amsmath, amsfonts, amssymb}
\usepackage{comment}
\usepackage{graphicx}
\usepackage{psfrag}
\usepackage[usenames,dvipsnames,svgnames,table]{xcolor}
\usepackage{enumerate}
\usepackage{jheppub}
\usepackage{soul}
\usepackage{subfig}

\newcommand{\be}{\begin{equation}}
\newcommand{\ee}{\end{equation}}

\def\be{\begin{equation}}
\def\ee{\end{equation}}
\def\bea{\begin{eqnarray}}
\def\eea{\end{eqnarray}}

\newcommand{\ep}{\epsilon}

\newcommand{\sig}{\sigma}

\renewcommand{\[}{\left[}
\renewcommand{\]}{\right]}
\renewcommand{\(}{\left(}
\renewcommand{\)}{\right)}
\newcommand{\w}{\wedge}
\newcommand{\vol}{\text{vol}}
\newcommand{\f}[2]{\frac{#1}{#2}}

\newcommand{\e}{\textrm{e}}

\renewcommand{\O}{\mathcal{O}}

\newcommand{\dd}{\mathrm{d}}

\renewcommand{\O}{\mathcal{O}}

\def\simleq{\; \raise0.3ex\hbox{$<$\kern-0.75em
      \raise-1.1ex\hbox{$\sim$}}\; }
   \def\simgeq{\; \raise0.3ex\hbox{$>$\kern-0.75em
      \raise-1.1ex\hbox{$\sim$}}\; }
      
      \newcommand{\figref}[1]{figure \ref{#1}}

\newcommand{\secref}[1]{section \ref{#1}}

\title{Is inflation from unwinding fluxes IIB?}
\author[\bigstar,\heartsuit]{Fridrik Freyr Gautason}
\author[\bigstar]{Marjorie Schillo,}
\author[\bigstar]{and Thomas Van Riet}

\emailAdd{ffg, thomas.vanriet, marjorie.schillo@kuleuven.be}

\affiliation[\bigstar]{\it Instituut voor Theoretische Fysica, K.U.Leuven,\\Celestijnenlaan 200D, B-3001 Leuven, Belgium  }
\affiliation[\heartsuit]{\it Institut de Physique Th\'eorique, Universit\'e Paris Saclay, CEA, CNRS,\\
Orme des Merisiers, F-91191 Gif-sur-Yvette, France }

\begin{document}

\abstract{ In this paper we argue that the mechanism of unwinding inflation is naturally present in warped compactifications of type IIB string theory with local throats. The unwinding of flux is caused by its annihilation against branes. The resulting inflaton potential is linear with periodic modulations. We initiate an analysis of the inflationary dynamics and cosmological observables, which are highly constrained by moduli stabilization.  For the simplified model of single-K\"ahler Calabi-Yau spaces we find that many, though not all of the consistency constraints can be satisfied.  Particularly, in this simple model geometric constraints are in tension with obtaining the observed amplitude of the scalar power spectrum.  However, we do find 60 efolds of inflation with a trans-Planckian field excursion which offers the hope that slightly more complicated models can lead to a fully consistent explicit construction of large field inflation of this kind. }

\maketitle

\section{Introduction}
The success of inflationary cosmology in describing the observed cosmic microwave background (CMB) has led to myriad slow-roll models constructed from an effective field theory point of view.  However, there are  various challenges in describing inflation via effective field theory \cite{Baumann:2014nda}. Further progress in inflationary cosmology  requires an understanding of inflation in a UV complete theory of gravity. In this paper we present a mechanism for inflation in string theory that can take place in a standard arena for string phenomenology based on type IIB flux compacifications using warped Calabi-Yau spaces with three-form fluxes \cite{Giddings:2001yu, Kachru:2003aw, Balasubramanian:2005zx, Westphal:2006tn}. This work stems from an investigation of whether the mechanism for \emph{unwinding inflation} \cite{D'Amico:2012sz, D'Amico:2012ji} can be embedded into the type IIB context and if so, to what extent must it be modified.

Unwinding inflation is based on the observation \cite{Kleban:2011cs} that Brown-Bunster bubbles \cite{Brown:1988kg} can be localized inside compact cycles, in which case they cross over the cycle periodically as they expand. In this way, a single instanton event can discharge many units of flux as the bubble moves over the periodic domain. This discharge  lowers the positive energy stored in the flux and may generate 60 efolds of inflation \cite{D'Amico:2012sz, D'Amico:2012ji}. Since this mechanism features universal ingredients of string theory (fluxes, branes and extra dimensions) it could lead to a natural model for inflation.  Furthermore, this mechanism has the potential to produce large field inflation, which is notoriously difficult to achieve in string theory.

In unwinding inflation there is a single $(p+2)$-form flux present\footnote{This could equivalently be taken to be the dual $(8-p)$-form.}, which is discharged by a  $p$-brane, but the flux backgrounds of \cite{Giddings:2001yu, Kachru:2003aw, Balasubramanian:2005zx, Westphal:2006tn} have multiple fluxes turned on presenting some complications to this basic mechanism.  Most notably, in the presence of multiple fluxes, one finds \emph{tadpole} conditions that require changes in flux quanta to be accompanied by changes in the net number of brane charges.  Particularly, in type IIB the three-form fluxes, $H_3$ and $F_3$, induce three-brane charge as can be derived from the Bianchi-identity for the five-form field strength:
\be \label{f5bianchi}
\dd F_5 = H_3\wedge F_3 + Q \delta\,,
\ee
where $Q\delta$ describes the localized three-brane charge density. Due to Gauss' law, the integral of this equation over a compact cycle must be zero. Hence, any change in the three-form fluxes must be accompanied by the creation or annihilation of three-branes. In other words, if inflation  features the decrease of either $F_3$ or $H_3$ via a five-brane bubble, the number of three-branes must change across the bubble wall.

The mechanism of \emph{brane-flux annihilation} \cite{Kachru:2002gs} provides a process whereby one can reduce flux quanta and the four-dimensional energy density within a controlled set of approximations. This process begins when anti-D3 branes are introduced to a background containing three-form flux.  Since the anti-D3 branes carry charge opposite to the charge induced by the three-form fluxes, they can annihilate such that both are reduced in a way that satisfies the tadpole condition.  Furthermore, as long as the antibranes can be treated as probes, so that the geometry can be argued to remain sufficiently close to a warped Calabi-Yau,  the antibranes induce a positive energy which is equal to twice their tension \cite{Maldacena:2001pb}. Therefore, as the flux \emph{and} the anti-D3 charge decreases together, so does the positive energy. This decrease in energy can be equivalently regarded as coming from a decrease in the $|F_3|^2$ contribution or from a decrease in the anti-D3 brane tension. Once the anti-D3 charges are annihilated the process can come to an end.

Anti-D3 branes and their possible annihilation with surrounding fluxes has been the subject of intensive study in the context of string cosmology.  Famously, KKLT \cite{Kachru:2003aw} argued that in the presence of a small number of anti-D3 branes there can be a meta-stable de Sitter state with a tunably small cosmological constant. Shortly thereafter, KKLMMT \cite{Kachru:2003sx}\footnote{See also \cite{Pajer:2008uy, Becker:2007ui} for related work.} embedded \emph{brane inflation} \cite{Dvali:1998pa}, in the KKLT background.  This model makes use of the attractive potential between  a mobile D3-brane and the aforementioned  anti-D3 branes to give rise to a sufficient period of inflation. These scenarios require that the string-scale energy of the anti-D3 branes  is sufficiently redshifted relative to the scale of moduli stabilization.  This gravitational redshifting occurs   inside \emph{throats}, i.e. regions of large warping  that act as  gravitational attractors for anti-D3 branes. We will follow the standard practice of modeling such throat regions with the Klebanov-Strassler (KS) solution \cite{Klebanov:2000hb}. 

The dynamical mechanism that we are interested in takes place in the same set-up as KKLMMT, but in a different region of parameter space.  We identify three regions in the parameter space spanned by the number $M$ of $F_3$ quanta threading the $S^3$ at the bottom of the KS throat, and the number $p$ of anti-D3 branes.  The three regions are
\be \label{regions}
\text{I.} ~~\,\,\frac{p}{M}< 0.08\quad\qquad \text{II.} ~~\,\,\frac{p}{M} \sim 0.08\quad\qquad  \text{III.}~~\,\, \frac{p}{M}\gg 1\,.
\ee
The KKLMMT model lives in region I where the KKLT vacuum is believed to exist.  A mechanism closely related to the one we propose, the \emph{giant inflaton} \cite{DeWolfe:2004qx}, is possible in region II, where the potential barrier against brane-flux annihilation turns into a shallow plateau.  The resulting potential can generate inflation via brane-flux annihilation\footnote{Although, the authors of \cite{DeWolfe:2004qx} conclude that it is not possible to get the requisite 60 efolds within the validity of their approximations.}. As we will show, unwinding inflation can be found in region III.  The unwinding mechanism we propose makes use of a flux cascade arising from the repeated brane-flux annihilation of anti-D3 branes  which are confined to the bottom of the throat.\footnote{A study of non-perturbative brane-flux annihilation in the KKLT setup was first carried out in \cite{Frey:2003dm}.} The role of the inflaton field in both unwinding inflation (region III) and the giant inflaton (region II) is played by the position of a \emph{fuzzy} NS5-brane which wraps the contractible $S^2$ on the $S^3$ at the bottom of the throat.  This fuzzy brane is the result of the anti-D3 branes polarizing in the flux background via the Myers effect \cite{Myers:1999ps}. The unwinding mechanism corresponds to the periodic motion of the NS5 moving back and forth from the north pole to the south pole of the $S^3$.  

One might expect that the limit of large $p/M$ is problematic because a large number of antibranes may produce a strong backreaction on the geometry. However, it is possible to retain the limit in which the size of the three-cycle, $R^2_{S^3} = \ell_s^2 g_s M$, is much larger than the radius of anti-D3 brane backreaction given by $R^2_{D3}=\ell_s^2 \sqrt{g_sp}$.  This only requires
\be
\frac{p}{g_s M^2} \ll 1~,
\ee
which is compatible with $p/M\gg1$. In \secref{FluxBack} we will also consider the effect of the antibranes and the flux cascade on the complex structure and K\"ahler moduli.  Using the simplified model of KKLT in which non-perturbative corrections are used to stabilize a single K\"ahler modulus, we are able to achieve a sufficient period of inflation. However, because in this set-up the antibranes provide the energy which uplifts the supersymmetric AdS$_4$ vacuum to de Sitter, we see that if all of the antibranes annihilate against flux the cascade will end in a vacuum with negative cosmological constant.  In \secref{discuss} we briefly discuss some possible dissipative effects that could serve to stop the cascade before all of the antibrane charge is gone.

Setting aside questions of reheating and focusing only on the period of 60 efolds, we find that in the KS throat the curvature of the $S^3$ leads to large oscillations in the second slow roll parameter, $\eta$. While the first slow roll parameter, $\epsilon$ remains small, these oscillations are translated into the power spectrum. A priori, large oscillations in the power spectrum are not incompatible with the observed CMB as long as their frequency is large enough. However, these oscillations complicate the use of the usual slow roll techniques and we are forced to solve the system numerically.  Initial investigations show a tension between fixing the correct amplitude of the power spectrum and satisfying all geometrical constraints.  A more complete study of parameter space is necessary in order to find an acceptable realization of the power spectrum, or robustly  rule out this version of our mechanism.   

In order to avoid  these issues, in \secref{torussec} we discuss our mechanism in a more speculative background where the cycle at the tip of a throat is a torus. This background is speculative because we do not know of an explicit example of a geometry that fits this description, however there are explicit examples of compact Calabi-Yau manifolds that contain three-cycles that are topologically tori (e.g. \cite{1999math......2076B}).   In this case the same process works using a  D5/anti-D5 pair (or NS5/anti-NS5 pair) moving periodically over a one-cycle in the torus.  Because the torus is flat, the troublesome oscillations can be made small allowing for standard slow roll inflation and agreement with CMB observations. 

Of particular importance is the fact that this mechanism  naturally allows for a trans-Planckian inflaton field range.  This occurs because the inflaton is identified with the position of a five-brane which moves repeatedly over the same fundamental domain of a compact cycle, and in each pass fluxes are annihilated against antibrane charges.  Therefore, there is no physical obstruction to achieving a large field range.  The monodromy effect is similar to axion monodromy inflation \cite{Silverstein:2008sg, McAllister:2008hb} (see also \cite{Marchesano:2014mla, Hebecker:2014eua, Escobar:2015ckf}): the periodic brane position is unfolded by the change in charge and energy. The resemblance to axion monodromy  extends to the effective potentials which are linear plus oscillations. A further discussion of the relation to the models of \cite{Silverstein:2008sg, McAllister:2008hb}, and particularly how the present model differs, is contained in \secref{discuss}.

Finally, we note that for the de Sitter vacuum of KKLT, or the inflationary scenario of KKLMMT to be valid, the anti-D3 branes have to be protected against direct brane-flux annihilation by a potential barrier.  The reliability of this potential barrier is currently under debate\footnote{See the following biased selection of recent papers \cite{Bena:2014jaa, Michel:2014lva, Cohen-Maldonado:2015ssa, Bena:2016fqp, Danielsson:2016cit} and references therein.}.  One of the primary objections to these scenarios is the use of the probe NS5 brane action at weak coupling \cite{Kachru:2002gs}.  This action is obtained by S-dualizing the D5 brane action and is strictly only valid at strong coupling.  In this paper we avoid this issue by placing our mechanism in the S-dual of KS at weak coupling. The unwinding process process is then mediated by a D5-brane, rather than an NS5, moving many times over an $S^3$ in the S-dual of the KS throat.   This may ameliorate some of the concerns regarding NS5 backreaction, and futhermore, in \secref{discuss} we will argue that the antibrane backreaction is expected to improve the agreement  with the CMB spectrum.

\section{The flux background} \label{FluxBack}
In this section we review the technical details of the flux background and discuss under what approximations the background remains stable during the cascade.  Because we are ultimately interested in an inflationary solution that could describe our universe, we need to begin with a compactification that exhibits a separation of scales between the compact directions and the length scales accessible to a four--dimensional observer.  We work in the well-studied type IIB supergravity compactifications of \cite{Giddings:2001yu} (see also \cite{Becker:1996gj,Dasgupta:1999ss,Greene:2000gh,Grana:2000jj} for related earlier work) where the ten--dimensional geometry is a warped product of a four--dimensional spacetime and a six--dimensional conformal Calabi-Yau manifold $X$. Denoting the Calabi-Yau metric by $g_{mn}$ we write the full metric as 
\be\label{backgroundmetric}
\dd s^2 = \ell_s^2\left(\e^{2A} \dd s_4^2 + \e^{-2A} {g}_{mn}\dd y^m\dd y^n\right)~.
\ee
The warp factor $\e^{A}$ only depends on the internal coordinates $y^m$, and $\dd s_4^2$ denotes the metric on the four--dimensional spacetime. The compactification on $g_{mn}$ leads to an effective ${\cal N}=1$ supergravity theory in four dimensions which is specified by the K{\"a}hler potential ${\cal K}$ and the superpotential ${\cal W}$. The tree-level superpotential is given by \cite{Gukov:1999ya}
\be\label{GVW}
{\cal W} = \int_X G_3 \w \Omega~,
\ee 
where $\Omega$ is the holomorphic $(3,0)$-form on $X$ and $G_3$ is the type IIB complex  three-form 
\be
G_3 = F_3 - \tau H_3~.
\ee
Here $F_3$ and $H_3$ are the RR and NSNS three-forms respectively and the axio-dilaton $\tau$ is defined by
\be
\tau = C_0 + i \e^{-\phi}~.
\ee
The three-form fluxes  give rise to masses for the many complex-structure moduli of $X$. These fluxes satisfy a quantisation condition, which in our convention takes the form
\be \label{fluxquant}
M_i\equiv\f{1}{(2\pi \ell_s)^2}\int_{\Sigma_i} F_3  \in {\bf Z}~,\qquad K_i\equiv -\f{1}{(2\pi \ell_s)^2}\int_{\tilde \Sigma_i} H_3 \in {\bf Z}~,
\ee
where the integrals run over a three-cycle $\Sigma_i$ and its Poincar\'e dual $\tilde \Sigma_i$. Each complex structure modulus -- which is roughly associated with a three-cycle -- receives a mass associated with the flux that is threaded on the Poincar\'e dual cycle. The superpotential \eqref{GVW} also provides a stabilization mechanism for the axio-dilaton through the appearance of $\tau$ in $G_3$. 

We are interested in an inflationary scenario that discharges some flux to gradually lower the four-dimensional vacuum energy; a natural candidate is one of the three-form fluxes, either $F_3$ or $H_3$. However, it is important that discharging such a flux does not upset the stabilization of the complex structure moduli.  Therefore, we will require that the amount of flux discharged is small compared to the total number of flux units.

As discussed above, these fluxes enter the right hand side of the Bianchi identity \eqref{f5bianchi} on the same footing as standard D3-branes.  Integrating the Bianchi identity over the compact manifold $X$ leads to the tadpole cancellation condition
\be\label{tadpole}
\f{1}{ (2 \pi \ell_s)^4 } \int_X H_3\w F_3 + N_\text{D3} = \f{\chi}{24}~,
\ee
where $N_\text{D3}$ counts the total quantized D3 brane charge and $\chi$ accounts for D3-charges of 7-branes in F-theory compactifications and is given by the Euler number of the F-theory fourfold \cite{Giddings:2001yu}. To accommodate the tadpole condition while discharging flux in the compactifications of \cite{Giddings:2001yu} (which include non-zero $H_3$-flux) we will employ the brane-flux annihilation mechanism of \cite{Kachru:2002gs}. In \secref{sec:d5onS3} we will explain the details of this mechanism, and how it can lead to a flux cascade giving rise to 60 efolds of inflation.  The cascade simultaneously decreases the number of units $F_3$ flux\footnote{We are currently discussing the mechanism in the S-dual of KS where $F_3$ flux is discharged and the regions described in \eqref{regions} are characterized by the value of $p/K$.}, $M$, and the number of antibranes present which we denote by $p$.  For each unit of $F_3$ that is discharged, the number of antibranes decreases by $K$, such that the tadpole condition is satisfied. The change in $M$ throughout the cascade will then be given by $p/K$, thus in order not to upset the complex structure stabilization we  require:
\be
p< KM~.
\ee

In addition to the complex structure moduli, we must ensure that our mechanism does not upset the stability of the K{\"a}hler moduli, which are not stabilized by fluxes. One can stabilize the K{\"a}hler moduli via non-perturbative quantum corrections \cite{Kachru:2003aw} or a combination of perturbative and non-perturbative as in \cite{Balasubramanian:2005zx}.  In this paper we use the simple example of a single modulus, $\rho$, stabilized by non-perturbative effects as in \cite{Kachru:2003aw}. The non-perturbative effects give rise to a correction to the superpotential
\be
{\cal W} = {\cal W}_0 + A_K\e^{i a_K\rho}~,
\ee
where ${\cal W}_0$ is given by \eqref{GVW}.  The modified superpotential leads to a nontrivial potential for the K{\"a}hler modulus \cite{Kachru:2003aw,Kachru:2003sx}
\be \label{sigpot}
V_K = \f{a_K A_K \e^{-a_K\sigma}}{2\sigma^2}\left(\f13\sigma a_K A_K \e^{-a_K\sigma} + {\cal W}_0 + A_K\e^{-a_K\sigma}\right)+\f{z^{4/3}}{g_s^2\sigma^2}{2p\mu_3 \over g_s}~,
\ee
where $\sigma=\text{Im}\rho$ and $z$ is the redshift factor discussed below.
When no antibranes are present, the potential has a minimum in which the moduli are stabilized in a supersymmetric AdS$_4$ vacuum. Including $p$ anti-D3 branes  provides the well-known uplift effect that can raise the vacuum energy density to positive values \cite{Kachru:2003aw}. In the $p\gg K$ regime of parameter space the antibranes are not stable and their decay gives rise to inflationary dynamics. This will result in $2p\mu_3/g_s$ in the last term  of \eqref{sigpot} being replaced by a function that depends on the position of the inflaton.  

Ensuring the stability of the  K{\"a}hler moduli throughout inflation constitutes one of the main challenges for any proposal for inflation in string theory (see e.g. the discussion in  \cite{Kachru:2003sx}.) Since we are interested in a large number of antibranes annihilating against many units of flux to ensure a long-lasting cascade, it would seem that stability is severely compromised. However, by placing the branes inside a deep warped throat where $z$ is small, the energy of the antibranes can be redshifted to a small value such that all geometric moduli remain stable throughout the process.  This constraint is in tension with an arbitrarily large inflaton field range, however in the example discussed in \secref{numsol} by a delicate tuning of the parameters in \eqref{sigpot} we achieve a trans-Planckian field range and  60 efolds of inflation in a controlled setting.

\subsection{The Klebanov-Strassler throat and its S-dual}
As noted above, it is desirable that the brane-flux annihilation be contained in a highly warped region of the internal manifold. Since the branes carry anti-D3 brane charge, they will naturally be attracted to such regions and the dynamics will therefore be confined in warped throats. We devote this section to a summary of a commonly used local representation of such a throat: the Klebanov-Strassler (KS) solution \cite{Klebanov:2000hb}. We also briefly discuss the S-dual of the KS solution which is used throughout \secref{sec:d5onS3}.

The KS solution is a non-compact example of a background that fits into the description of \cite{Giddings:2001yu,Grana:2000jj}.  In the throat region the type IIB axio-dilaton is constant 
\be
\tau = \f{i}{g_s}~.
\ee
In the bulk of the Calabi-Yau, this need not be true: the presence of seven-branes wrapping cycles of the internal manifolds  leads to dynamical axio-dilaton. In this case the proper framework to describe the background is F-theory. However, these details will be unimportant for our purposes as we are only interested in dynamics taking place deep in the throat where the dilaton is constant. The remaining type IIB supergravity fields satisfy the equations
\bea
{\e^{4A} g_s^{-1}} &=& \alpha~, \\
\star_6 G_3 &=& i G_3~,\label{ISD}
\eea
in the gauge 
\be \label{C4gauge}
{C}_4 = \alpha \,\vol_4~,
\ee
where $\vol_4$ is the volume form of $\dd s_4^2$. 

The KS solution describes a  deformation of the singular \emph{conifold} \cite{Candelas:1989js}:
\be\label{conifoldcone}
\dd s_6^2 = \dd r^2 + r^2 \dd s_{T^{1,1}}^2~,
\ee
where $\dd s_{T^{1,1}}^2$ is the metric on the Sasaki-Einstein manifold $T^{1,1}$, which is topologically $S^2\times S^3$. In the presence of three-form fluxes the deformation of the conifold replaces the singular region of the conifold metric ($r\to 0$) with a smooth space by blowing up the $S^3$ at the tip to a finite size. In this tip region the metric takes the form
\be\label{conifoldtip}
\dd s_6^2 \to \dd \tilde r^2 + \dd\Omega_3^2 + \tilde r^2 \dd \Omega_2^2~.
\ee
The full type IIB solution on the deformed conifold with a metric that interpolates between the tip region \eqref{conifoldtip} and the cone region \eqref{conifoldcone} is known \cite{Klebanov:2000hb} but we will not require its precise form as the antibrane dynamics are confined to the tip region. The warp factor in the tip region is constant and fixed to be:
\be \label{KStip}
\e^{-2A_\text{tip}} \simeq M g_s~,
\ee
where henceforth $M$ refers to the  $F_3$ threaded through the three-sphere at the tip.

The KS solution can be embedded in compact space by sewing it to a compact Calabi-Yau. The description of the throat breaks down and the bulk Calabi-Yau description takes over when $\e^{2A}$ reaches the value determined by the hierarchy between the tip and the bulk. This hierarchy was calculated in \cite{Giddings:2001yu} and is
\be \label{KShierarchy}
\e^{2(A_\text{tip}-A_\text{bulk})} \sim z^{2/3} \sim (M g_s)^{-1}\e^{-\f{4\pi K}{3 M g_s}}~,
\ee
where $K$ is the number of units of $H_3$ that must have legs along the $\tilde r$ and $S^2$ directions perpendicular to the $S^3$ such that the imaginary self-dual (ISD) condition \eqref{ISD}  is satisfied.

A very similar background can be obtained by S-dualizing the KS solution (SDKS). Since the dilaton in \cite{Klebanov:2000hb} is a modulus we can dial it to large values, perform the $SL(2)$ transformation and end up with a weakly coupled background. The physical difference between KS and its S-dual is therefore only in the fluxes. The KS solution has $M$ units of $F_3$ flux threading the $S^3$ at the tip whereas SDKS has $K$ units of $H_3$-flux at the tip. This will be important for us as the difference will result in a flux cascade involving an NS5 brane in KS \cite{Kachru:2002gs} or D5 branes in SDKS. We expect that both throat backgrounds should be common in the landscape of type IIB compactifications, and so we discuss both possibilities. While the behavior of the metrics in both solutions is virtually identical, the different role of the three-form fluxes translates into different expressions for \eqref{KStip} and \eqref{KShierarchy}
\bea
\text{SDKS:}& \qquad \e^{-2A_\text{tip}} \simeq K~, \label{SDKStip}\\
&  \qquad \e^{2(A_\text{tip}-A_\text{bulk})} \sim K^{-1}\e^{-\f{4\pi M g_s}{3 K}}~. \label{SDKShierarchy}
\eea

Since the flux cascade is confined to the tip region, we must make sure that the branes that mediate the cascade do not destabilize the local geometry of the tip. This simply means keeping the horizon radius of the antibranes small compared to the local geometry of the tip. The size of the tip geometry is set by \eqref{KStip} and \eqref{SDKStip} in KS and SDKS respectively whereas the horizon radius of the antibranes is determined by $g_s p$.  Therefore, the probe approximation will be valid as long as the following are satisfied:
\bea
\text{KS:}& \qquad  R^2_{S^3}=\ell_s^2M g_s  \gg \ell_s^2\sqrt{g_s p}=R_\text{D3}^2~, \label{KSprobe}\\
\text{SDKS:}&  \qquad R^2_{S^3}=\ell_s^2 K \gg \ell_s^2 \sqrt{g_s p}=R_\text{D3}^2~. \label{SDKSprobe}
\eea

\section{Inflation from cascading brane-flux annihilation}\label{sec:d5onS3}
\begin{figure}
\begin{center}
\includegraphics[width=.3\textwidth]{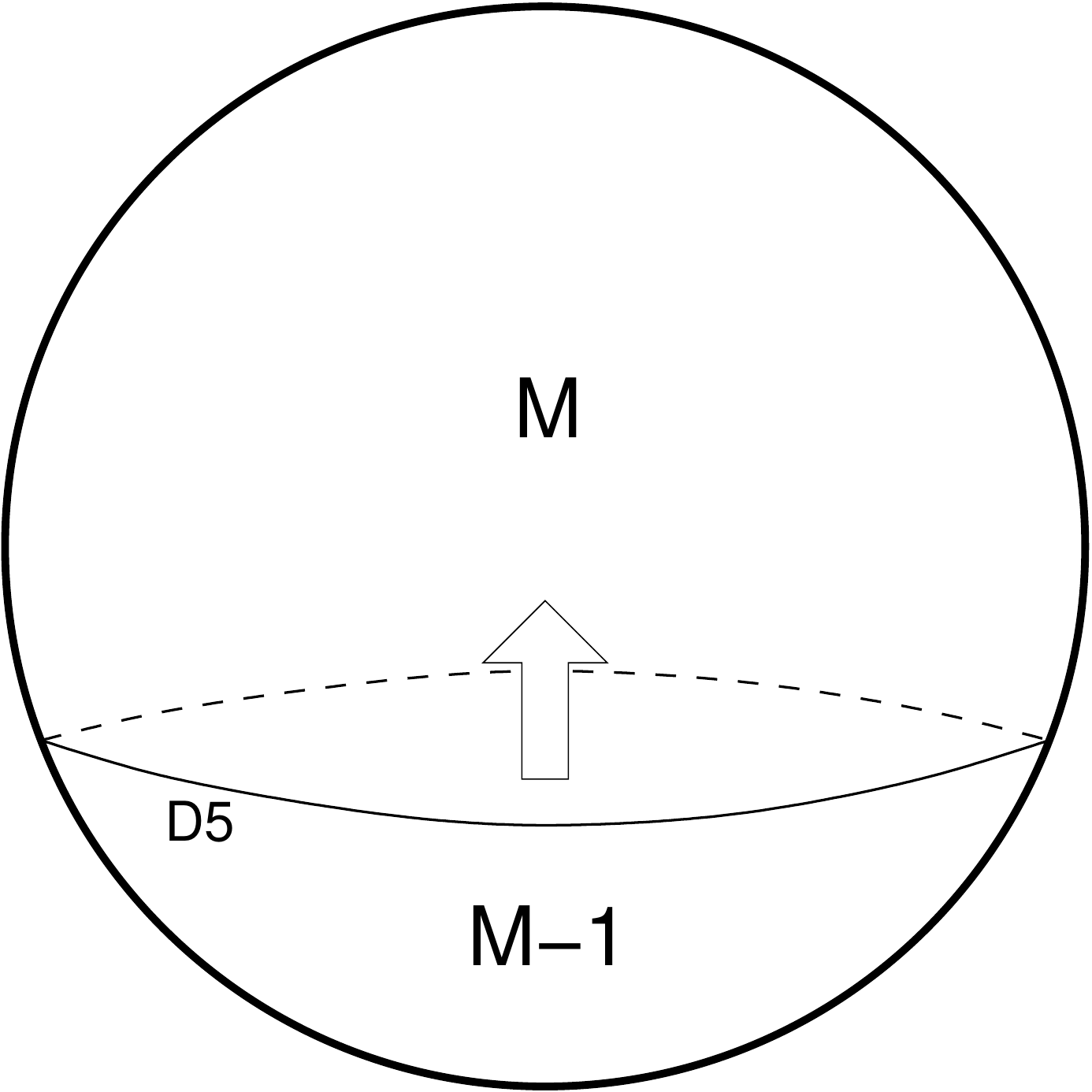}
\caption{Schematic representation of the polarized five-brane on an $S^3$.  The flux cascade corresponds to the periodic motion of the brane between the two poles. \label{spherebrane}}
\end{center}
\end{figure}

In this section we  describe in detail how the brane-flux annihilation of \cite{Kachru:2002gs} proceeds when a large number, $p\gg1$, of anti-D3 branes are placed into a throat region.  We will see that in this case the five-brane must pass over the sphere many times before reaching the supersymmetric vacuum.
In contrast to \cite{Kachru:2002gs} we begin SDKS where the anti-D3 branes polarize into a D5-brane that wraps an $S^2$ inside the $S^3$ at the tip of the throat (c.f. \figref{spherebrane}). This three-sphere carries $K$ units of $H_3$-flux, and as the D5 moves in the $S^3$,  $F_3$ flux is discharged in the dual cycle. 

Througout this section we will only be interested in the tip region of the throat where the antibranes are confined. The metric there takes the local form
\be \label{tipmetric}
\dd s^2 = \ell_s^2\left(\e^{2A} \dd s_4^2 + \e^{-2A} \left(\dd \psi^2 +\sin^2(\psi)~ \dd\Omega_2^2 + \dd s_{M_3}^2 \right)\right)~.
\ee
For now, we will continue using the generic warp factor $\e^{2A}$, as opposed to restricting to the value fixed by the deformation of the conifold given in \eqref{KStip} or \eqref{SDKStip}.

\subsection{The action} \label{theaction}
We start with the probe action of a D5:
\be \label{D5S3}
S = {-\mu_5 \over g_s} \int d^6 \xi \[-\det(G_\parallel)\det(G_\perp -  {\cal F}_2 )\]^{1/2} -\mu_5 \int \left\{C_6 + {\cal F}_2\w C_4\right\}~,
\ee
where $G_\perp$ is the induced metric along the $S^2$, $G_\parallel$ is the metric along the non-compact and $\psi$ directions, and ${\cal F}_2 =2 \pi \ell_s F_2 +B_2$.  To use this action to solve for the dynamics of the D5, we simply need to compute each component as a function of the position of the D5 in the compact space.  

Starting with the Chern-Simons action, we note that in the gauge chosen in \eqref{C4gauge}, 
\be
F_7 = -\star_{10} F_3 =   H\w C_4 ~.
\ee
Since $F_7 = \dd C_6+H\w C_4$, this implies that $C_6$ is pure gauge and can be set to zero.  The other term in the Chern-Simons action, $\int {\cal F}_2 \w C_4$, is the coupling that allows the D5-brane to carry D3 charge.  Schematically,  whatever sits in front of $C_4$ is the effective D3 charge -- therefore at the beginning of the cascade this should be $-\mu_3 p$.   Looking at ${\cal F}_2$ we see that because $K$ units of $H_3$-flux thread the three-sphere spanned by $\psi$, flux quantization \eqref{fluxquant} gives:
\be
B_2 = -K\ell_s^2 \left(\psi-\f12 \sin(2\psi)\right) \vol_2 ~,
\ee
where $\vol_2$ is the volume form on $S^2$. At the beginning of the cascade $\psi \approx 0$ and this term vanishes.  This allows us to fix the world volume field strength, $F_2= {p \ell_s\over 2} \vol_2$. Integrating over the $S^2$ in the Chern-Simons action one can check
\be
Q_{\rm D3}=-\mu_5 \int_{S^2} {\cal F}_2  = -{(2\pi \ell_s)^2 \mu_5 K \over \pi}\,U(\psi)~.
\ee
Here, we have defined $U(\psi)$, which measures the D3 charge:
\be
U(\psi) =  \f{\pi p}{K}- \psi+\f12 \sin(2\psi)~.
\ee
Since $\mu_3=(2\pi\ell_s)^2\mu_5$, we see that when $\psi=0$ we start with the correct amount of anti-D3 charge, and this charge decreases by $K$ units each time $\psi$ increases by $\pi$.  The tadpole condition \eqref{tadpole} is satisfied by decreasing $M$ by one unit as the D5 passes across the $S^3$.  It is in this sense that the anti-D3 branes annihilate against the $F_3$ flux.  It is  clear that in order to achieve a flux cascade we will need:
\be
{p\over K} \gg 1~.
\ee

Next, we need to evaluate the kinetic term in \eqref{D5S3}.  As mentioned above, $G_\perp$ is simply the metric on the $S^2$ (c.f. eq. \eqref{tipmetric}.) Using the values for ${\cal F}_2$ from above we can write:
\be
\sqrt{\det(G_\perp -  {\cal F}_2 ) }= \sqrt{ \e^{-4A}\sin^4(\psi)+K^2 U^2(\psi)}\sqrt{g_{S^2}}~.
\ee
The metric $G_\parallel$ is the induced metric in the spacetime directions.  The D5 should be thought of as a bubble in the three extended spatial directions and the $\psi$ direction (while it trivially wraps the $S^2$.)  Then, neglecting perturbations, an observer at a fixed position in spacetime will see $\psi$ as a function of $t$ alone:
\be
\sqrt{-\det(G_\parallel) }= \ell_s^4\e^{4A}a^3(t)\sqrt{1-\e^{-4A}\dot{\psi}^2}~,
\ee
where $a(t)$ is the scale factor of a Friedmann-Lema\^itre-Robertson-Walker (FLRW) spacetime.

\begin{figure}[h]
\begin{center}
\includegraphics[width=.65\textwidth]{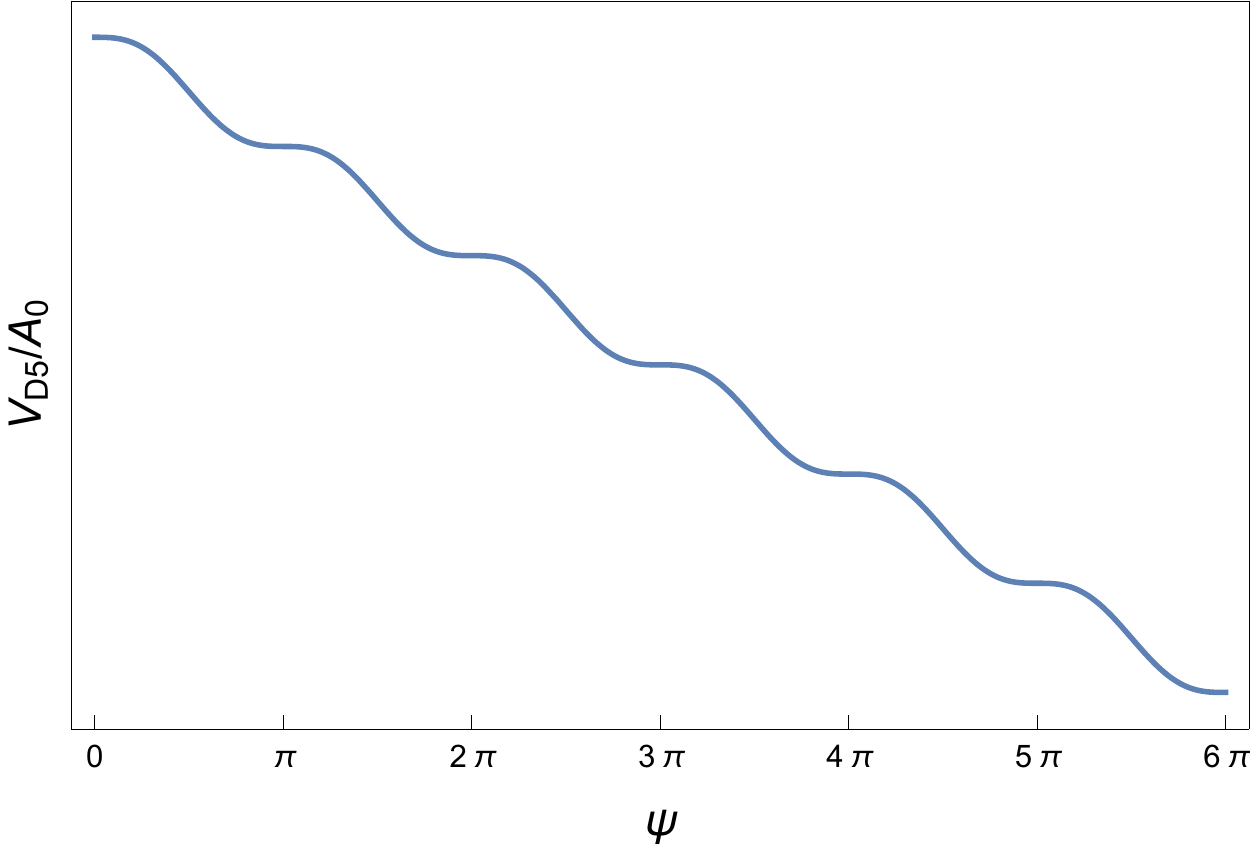}
\end{center}
\caption{\label{potcompare} The D5 potential neglecting the overall dimensionful factor. The only relevant parameters are the ratio $p/K$ (or $p/M$ if you consider the NS5), which we set to 50. Here we have specialized to the SDKS case where $\e^{-4A} = K^2$.}
\end{figure}

Combining the kinetic terms and the Chern-Simons action and integrating over the $S^2$ we write the four-dimensional action for the D5:
\be  \label{actionD5}
S = - A_0\int \dd^4  x \, a^3(t)\,\e^{4A}\(V_2(\psi) \sqrt{1-\ell_s^2\e^{-4A} \dot{\psi}^2} +U(\psi)\)~,
\ee
with 
\be
V_2(\psi)=\sqrt{\f{\e^{-4A}}{K^2}\sin^4(\psi) +  U(\psi)^2}~, \qquad A_0 =  {\mu_3  K \over g_s \pi}~.
\ee
In \eqref{actionD5} we have rescaled the four-dimensional coordinates by a factor of $\ell_s$ such that they are now dimensionful.  The position of the D5 in the $\psi$-direction will play the role of the inflaton.  Setting the inflaton kinetic energy to zero, we find the inflaton potential (\figref{potcompare}): 
\be\label{D5potential}
V_\text{D5}(\psi) = A_0\, \e^{4A}\left[\sqrt{\f{\e^{-4A}}{K^2}\sin^4(\psi) +  U(\psi)^2}+ U(\psi)\right]~.
\ee

In the following sections we will examine the dynamics of the cascading brane-flux annihilation that takes place  in region III, $p\gg K$ (c.f. \eqref{regions}). We are interested in seeing whether the resulting cascade process can give rise to inflationary dynamics, and furthermore, when inflation is possible, what are the CMB observables predicted by this model.  

\subsection{Regime of validity and comparison to NS5 in KS}

Before moving on to the inflationary dynamics, it is interesting to note what happens if we look at this process in KS, rather than its S-dual.  If we re-do the calculation of this section in the KS background, the resulting potential is the potential for an NS5 brane \cite{Kachru:2002gs}:
\be\label{KPVpotential}
\begin{split}
V_{\rm NS 5}(\psi) &= \f{\mu_3}{g_s \pi}  M\left[\sqrt{ \f{\e^{-4A}}{(M g_s)^2}\sin^4(\psi) + \tilde U(\psi)^2} + \tilde U(\psi)\right] ~, \\
\tilde U(\psi) &=  \f{\pi p}{M g_s}- \psi+\f12 \sin(2\psi)~.
\end{split}
\ee
The potentials in both cases are subject to the following string of inequalities:
\be
\begin{split}
{\rm KS}:& \quad (M g_s )^2  \gg g_s p \gg M g_s~, \\
{\rm SDKS}:& \quad K^2  \gg g_s p \gg g_s K~,
\end{split}
\ee
where the first inequality  follows from making the antibrane backreaction small compared to the size of the cycle, and the second inequality is the condition for having a cascade.  The cascade condition changes between KS and SDKS because in the S-dual case that we consider, we discharge the $F_3$-flux, whereas in the original case of \cite{Kachru:2002gs}, which takes place in KS, the $H_3$ flux is discharged.

\subsection{A canonically normalized approximation} \label{cannormsphere}
To solve the system described by \eqref{actionD5} it is necessary to use numerics, however in order to gain some intuition we will first make several approximations that allow us to canonically normalize the scalar field and make analytic estimates for the cosmological observables.  First, we expand the DBI kinetic term for small velocity, keeping terms up to second order, $\O\(\e^{-4A} \dot{\psi}^2\)$.  Then we expand in large $p/K$ - this is the quantity that counts the number of steps in the cascade.  In the second expansion we keep terms at next to leading order, which is $(p/K)^0$, but drop terms of order $(p/K)^0\dot{\psi}^2$ as they are also next to leading order in the velocity expansion.  These expansions will ultimately need to be justified by comparison to the full numerical solution, and we find that for a certain range of parameters they are appropriate.

These approximations allow us to write the action for a canonically normalized scalar, $\phi = \sqrt{\ell_s^2A_0 \pi p /K}\psi \equiv f \psi$:
\be \label{Scannonnorm}
S=\int d^4x a^3(t) \(\frac12 \dot{\phi}^2 - 2A_0\e^{4A}\(\frac{\pi p}{K}- {\phi \over  f} + \frac12 \sin\Big({2\phi \over 
f}\Big)\)-\Lambda\)~.
\ee
Here we have added the negative cosmological constant corresponding to the supersymmetric vacuum for the K\"ahler moduli prior to adding antibranes.  In the next subsection we will present a numerical solution that generates 60 efolds of inflation and $\Lambda$ is calculated via the  potential \eqref{sigpot}.

 We can  apply standard methods to calculate inflationary observables for a scalar field with a linear potential plus oscillations \eqref{Scannonnorm}. Particularly, we will be interested in finding at least 60 efolds of inflation that result in a power spectrum in agreement with the observed value.  These quantities are given by the standard formulae\footnote{The DBI kinetic term results in non-trivial speed of sound, $c_s$.  However, for all realizations of this model we find $c_s \sim 1$ and $c_s$ varies adiabatically such that ${\cal P}_\zeta$ in \eqref{standardeqn} is valid.} \cite{Garriga:1999vw}:
\be \label{standardeqn}
N = \int {H \over \dot{\phi}} d\phi \qquad {\cal P}_\zeta = {H^2 \over 8 \pi^2 M_{pl}^2 \ep} \qquad \ep = {\dot{H} \over H^2}~.
\ee

While the potential of \eqref{Scannonnorm} is monotonically decreasing, the oscillations are not small in that $\partial_\phi V(\phi =n\pi f)=0$, where $n$ is an integer.  The fact that the derivative of the potential goes to zero means that we cannot be in the traditional slow roll regime where one makes the approximation that acceleration is negligible and $\dot{\phi}\approx V' /3H$.  In this regime our inflaton gets stuck at the first pole where $\dot{\phi}\to 0$ and we do not see a cascade.  However, scenarios in which the second slow roll parameter, $ \eta=\dot{\ep}/(\ep H)$, is large are not ruled out as long as $\eta$ oscillates and $\ep$ remains small.  In these cases acceleration is not negligible and the inflaton will not get stuck.  

Despite these issues we will continue by dropping the oscillating term in the potential:
\be
V_\text{lin} = 2A_0\e^{4A}(\pi p/K-\phi/f)~.
\ee
This will obviously cause us to miss the oscillations in both the field velocity and also the resulting power spectrum, however we will be able to attain the average behavior, which is useful for the order of magnitude estimates we seek. Subtleties arising due to the oscillations, as well as cases that have large $\eta$ will be discussed in \secref{numsol} where we examine the full numerical solutions that these approximations are meant to capture. 

Using the simplified linear model and the slow roll approximations, 
\be
H_\text{lin} = \sqrt{V_\text{lin}/(3M_{pl}^2)} \quad \text{and} \quad \dot{\phi}_\text{lin} = -\partial_\phi V_\text{lin}/(3H_\text{lin})~,
\ee
 one can calculate the total number of efolds, as well as the value of $\phi$ that corresponds to 60 efolds before the end of inflation
\be
N_\text{tot}\simeq { f^2 \over 2 M_pl^2} \( {\pi p \over K} +{ \e^{-4A} \Lambda \over 2 A_0}\)^2  \qquad 
\phi_* = f\({ \pi p \over K} + {\e^{-4A}\Lambda \over 2A_0}\)-2\sqrt{30}M_{pl}~,
\ee
where the end of inflation is set to be the point where the potential energy reaches zero.   Given these assumptions, the power spectrum, ${\cal P}_\zeta$, is simply given by:
\be \label{pspeclin}
{\cal P}_\zeta|_{\phi_*} = {40\sqrt{30} A_0 \e^{4A}  \over  \pi^2  M_{pl}^3 f}~.
\ee
The challenge now is to determine whether there exists a set of parameters in which $N_\text{tot} \gtrsim 60$, ${\cal P}_\zeta|_{\phi_*} \sim 10^{-9}$ and we are within the regime of validity of the probe approximation and other requirements for a stabilized geometry.  

The four-dimensional Planck scale, $M_{pl}$, can be expressed in terms of the other parameters:  
\be 
M_{pl}^2 = 2 {\int d^6y \e^{-4A} \sqrt{g_6} \over (2\pi)^7 \ell_s^2 g_s^2}\equiv  2 {{\cal V} \over (2\pi)^7 \ell_s^2 g_s^2}~,
\ee
where $g_6$ is the unwarped metric and  the spacetime coordinates are dimensionful while the internal coordinates are not.
In writing the parametric dependence of $M_{pl}$, we introduced a new parameter, the warped volume of the Calabi-Yau denoted by ${\cal V}$.  We require that the total warped volume is greater than the warped volume of the throat region which  is:
\be \label{throatvol}
{\cal V}_{\text{throat}} \simeq \sqrt{27 \pi^9} (g_s K M)^{3/2}\e^{{4\pi M g_s  \over 3 K}}~.
\ee
The exponential factor in this expression results from the exponential hierarchy between the warp factor at the bottom of the throat and the bulk of the Calabi-Yau \eqref{SDKShierarchy}.  This volume is also computed for the KS throat in \cite{DeWolfe:2004qx}\footnote{The differing exponential factor here is due to the fact that we use warped units, whereas \cite{DeWolfe:2004qx} does not.}. 

One is left with a six-dimensional parameter space spanned by $p$, $g_s$, $M$, $K$, $\Lambda$ and ${\cal V}$.  One of these can be fixed by requiring that the power spectrum \eqref{pspeclin} to its observed value $\sim 10^{-9}$.  However, the remaining five-dimensional parameter space must satisfy our collection of constraints:
 \begin{itemize}
\item The probe approximations: $M,K\gg1$, $MK\gg p$, and $\sqrt{g_s p/K^2} \ll 1$
\item The inflaton is the only light scalar during inflation requires $H\ll M_{KK}={\cal V}^{-1/6}$, where $M_{KK}$ is the mass of the bulk Calabi-Yau Kaluza-Klein modes.  Additionally the masses of the infrared  modes, $z^{1/3}M_{KK}$, must be heavier than the Hubble scale. 
\item In order to  stabilize the geometry the magnitude of $\Lambda$ cannot be much less than the uplift energy from the antibranes:
\be
|\Lambda| \lesssim {2 \mu_3 p \over g_s K^2}~.
\ee 
\item The warped volume of the Calabi-Yau is larger than the warped volume of the throat:
\be 
{\cal V} >{\cal V}_{\rm throat}~.
\ee
Due to the large hierarchy required to redshift the energy of the antibranes, the throat volume \eqref{throatvol} must be large in string units. This means that the K{\"a}hler modulus $\sigma$ which is related to the unwarped volume should be stabilized at a large value. In this paper, we take the warped volume of the Calabi-Yau to be a free parameter and do not directly relate it to $\sigma$.  This is not strictly valid, however, due to the large warping the relationship between ${\cal V}$ and $\sig$ is non-trivial (see references \cite{Giddings:2005ff,Shiu:2008ry,Frey:2008xw,Martucci:2009sf}).  We leave a full computation of the relationship between these parameters for future work.
\item There are at least 60 efolds: $N_\text{tot}/60>1$.
\item The cascade occurs: $p/K>1$ and $\eta \gtrsim 1$.  Although the slow roll approximations push us into the regime of small $\eta$, if it is too small the brane gets stuck at the poles and there is no cascade.
\end{itemize}

It is possible to find examples which satisfy all of these constraints, however we have not been able to simultaneously satisfy all constraints and find an observationally valid power spectrum.  Until a systematic exploration of this high-dimensional, highly-constrained parameter space has been carried out, we cannot either rule out or accept this model.

 \subsection{Full numerical solutions} \label{numsol}
\begin{figure} 
\begin{center}
\includegraphics[width=.65\textwidth]{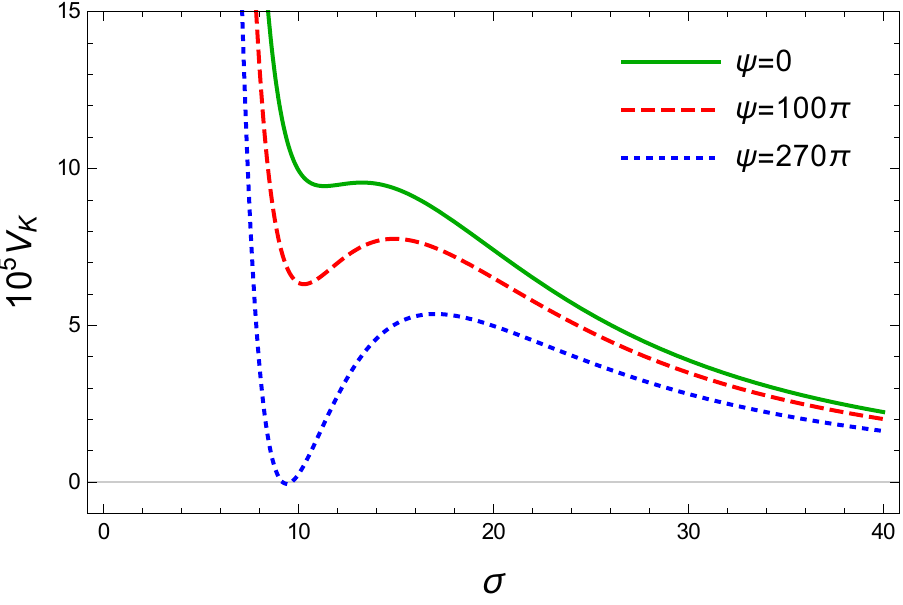} 
\caption{The evolution of the K\"ahler potential throughout the cascade.  It is important that the modulus is stable and that its value, $\sig_*$, does not evolve significantly.  Although the minimum of the potential at $\psi=0$ looks dangerously shallow, the reader should bear in mind that the $\psi-$direction of the potential is unstable.}
\label{Kahlerfig}
\end{center}
\end{figure}

In this subsection we will use the analysis of the previous subsection to find a set of parameters which satisfy all constraints.  Using these parameters and the intuition given by analytic estimates, we  take into account proper stabilization of the K\"ahler modulus and solve the system numerically.  We begin by considering the effect of the cascade on the potential for $\sig$ given in \eqref{sigpot} \cite{Kachru:2003aw,Kachru:2003sx}.  The uplift term is proportional to the potential energy of $p$ stationary anti-D3 branes, $2\mu_3p/g_s$.  During the flux cascade this potential energy is replaced by the potential energy \eqref{D5potential}:
\be
{2\mu_3p\over g_s} \to \e^{-4A} V_{\rm D5}(\psi) = K^2 V_{\rm D5}(\psi)~,
\ee
 where here and throughout the rest of this section we specialize to SDKS where $\e^{-4A} = K^2$.  Instead of solving the coupled system of the K\"ahler modulus  and the position of the D5 simultaneously, we will first find a stable potential for  $\sig$, and then check that its value does not evolve too much (or destabilize) throughout the cascade.  In \figref{Kahlerfig} we show the evolution of the K\"ahler potential throughout the cascade for the parameters given in table \ref{params}.  Using these values  we can read off the negative energy density in the supersymmetric vacuum and add this $\Lambda$ to the D5 brane action, \eqref{actionD5}.

\begin{figure} 
\begin{center}
\begin{tabular}{ c c }
\includegraphics[width=.48\textwidth]{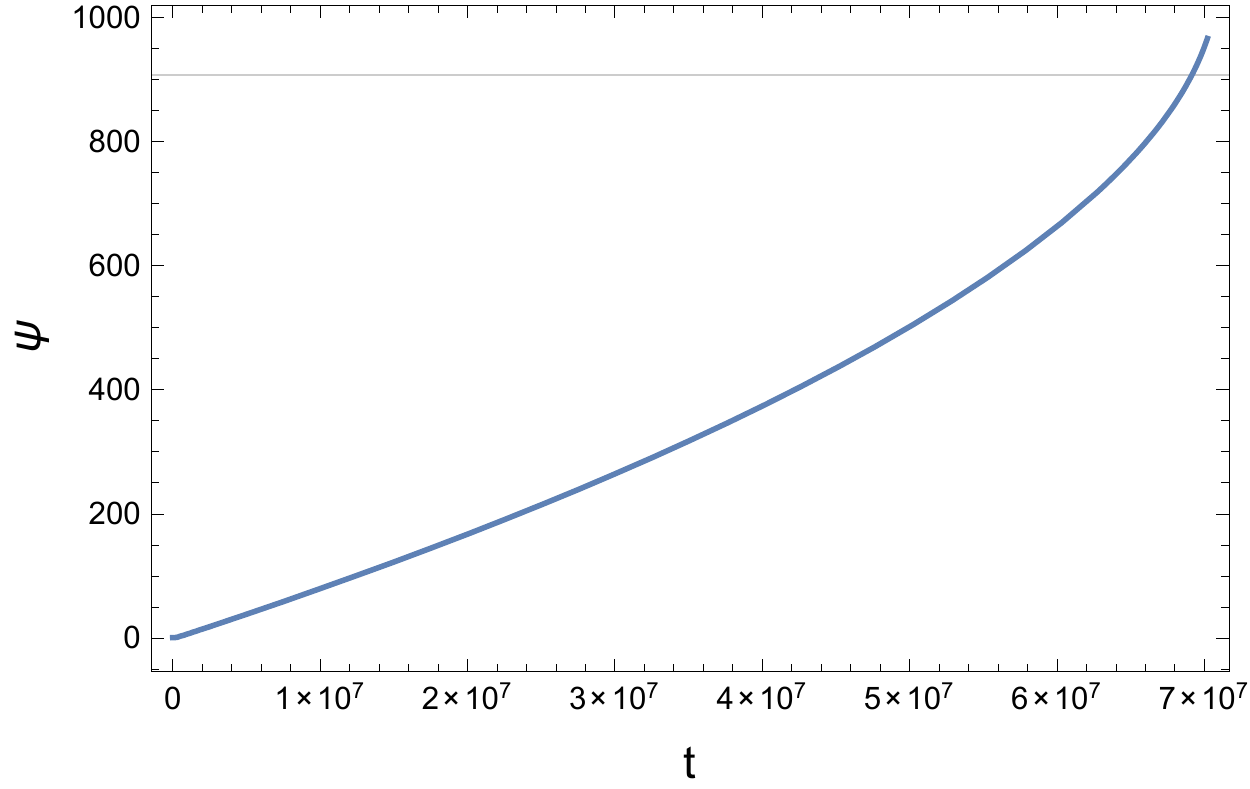} & \includegraphics[width=.48\textwidth]{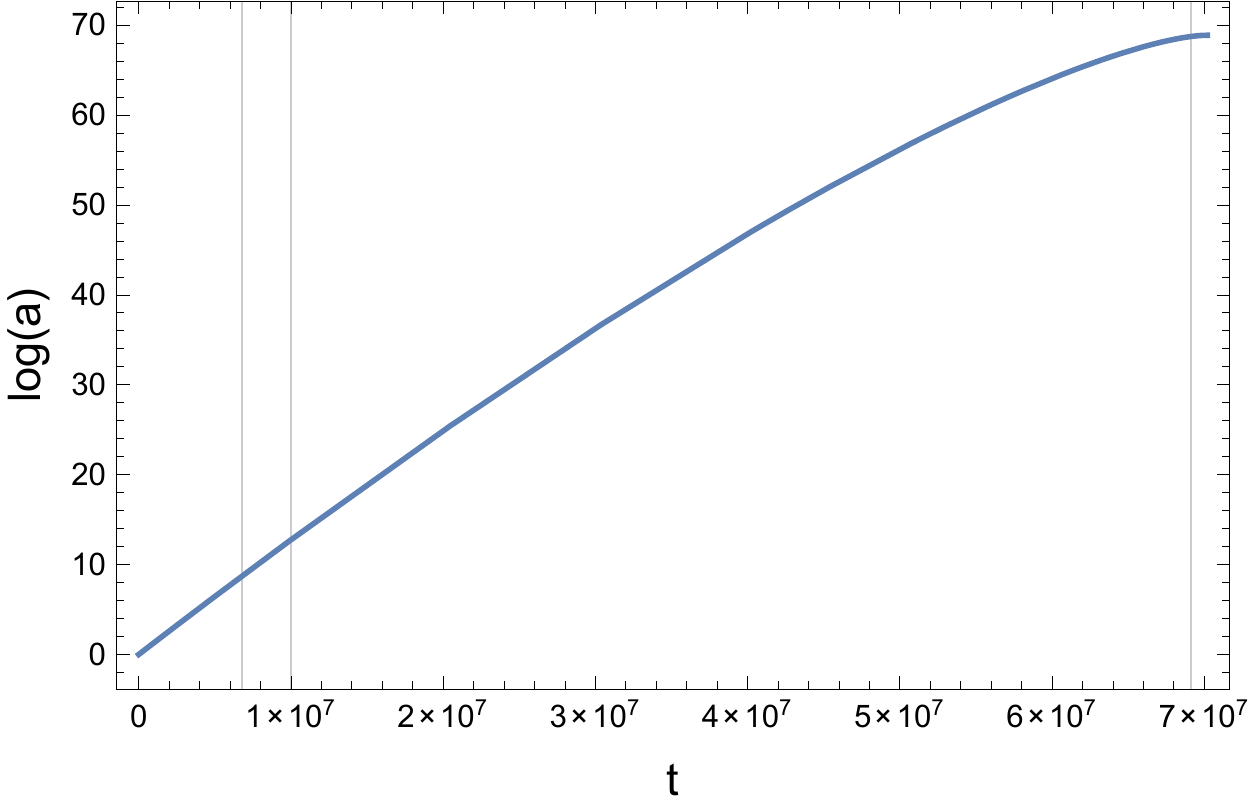}
\end{tabular}
\caption{\textbf{Left:} the position of the D5 as a function of time in the case of parameters given by table \ref{params}.  The horizontal line marks the end of inflation where the potential energy is zero. \textbf{Right:} The log of the scale factor with vertical lines showing the observational window and the end of inflation. The total period of inflation is $69$ efolds.}
\label{figs}
\end{center}
\end{figure}

Once the K\"ahler modulus is stabilized, we are ready to solve the inflationary dynamics. Following \cite{DeWolfe:2004qx} we pass to the Hamiltonian formalism and solve the system of first order equations.  We give the parameters and the degree to which they satisfy our constraints for a typical example in table \ref{params}, and show the resulting dynamics \figref{figs}.  There is an additional  caveat regarding these parameters. The product $KM$ at the end of inflation is ${\cal O}(10^6)$ and should be cancelled by $\chi/24$ in \eqref{tadpole}.  This implies that the Euler number of our fourfold is an order of magnitude larger than the largest known Euler number of an elliptically fibered  fourfold\footnote{We thank Liam McAllister and Alexander Westphal for pointing this out to us.}.  While it is not impossible that a Calabi-Yau with larger Euler number exists, it would be preferable to find a set of parameters with smaller $KM$.

\begin{table} 
\begin{tabular}{|c|c|c|c|}
\hline
 $\Delta \phi/M_{pl} = 12.1$ & $H/M_{pl}=  6.5\times 10^{-11} $ & $H/M_{KK}=  1.7\times 10^{-4}$ & ${\cal V} = 5.3 \times 10^{12}\ell_s^6$\phantom{\Big)}\\ \hline
$z^{1/3} = .012$ & ${\cal V}/{\cal V}_\text{throat} = 1.1$ & $g_s p /K^2 =.06 $ & $p/KM = .54$ \phantom{\Big)}\\ \hline
$p=4.5 \times 10^6$ &  $K=4500$  &  $M=1852$ & $g_s=.27\phantom{\Big)}$ \\ \hline
$A_K=3$ &  $a_K=2\pi/31$  &  ${\cal W}_0 = 1.31$ & $\sig_*=10.4 \phantom{\Big)}$\\ \hline
\end{tabular}
\caption{One set of parameters that satisfies our constraints. We have chosen the average value of $\sig_*$ throughout the cascade.}
\label{params}
\end{table}
As mentioned in the previous section, this example has large oscillations in the slow roll parameters which translate into large oscillations in the power spectrum.  Although the first slow roll parameter, $\ep = \dot{H}/H^2$, is always small, the second slow roll parameter, $\eta = \dot{\ep}/(\ep H)$ is oscillating with a large amplitude.  This is not the typical $\eta$-problem, where $\eta$ becomes large, drives $\ep$ to become large, and ends inflation before 60 efolds, however it is still a problem in that even oscillations in a small $\ep$ translate into large oscillations in the power spectrum.  However, it may be possible to find an acceptable set of parameters in which these oscillations are fast, i.e.  10 per efold.  In this case late time physics can smooth them out in agreement with observations.  These oscillations will also give rise to resonances and resonant non-Gaussianity \cite{Flauger:2009ab, Flauger:2010ja}.

The results presented here are not in agreement with the scale invariant spectrum that we observe.  The magnitude of the power spectrum is smaller than the observed value.  This is the reason that despite having an trans-Planckian field range, the scale of inflation is much lower than the GUT scale. However, we should stress that the parameter space for this model is far from being fully explored.  In the absence of analytic estimates, which were only available for a small portion of the parameter space, more sophisticated techniques must be employed to impose constraints and find acceptable power spectra.  We leave this to future work.  Despite the apparent difficulties in KS and SDKS, if one allows for less well-understood geometries, we are able to find observationally viable realizations.  This is the topic of the next section.

\section{Brane-flux annihilation on $T^3$} \label{torussec}
The large oscillations in the power spectrum that we found in the previous section can be traced to the curvature of the sphere.  Therefore, we expect that if the flux cascade takes place on an flat internal 3-manifold the power spectrum will not suffer from these large oscillations.  We will consider a cascade that takes place on $T_3$, whose coordinates $T_i$ are intervals from 0 to $L_{i}$.  Despite the fact that there are no one cycles within Calabi-Yaus, toric special Lagrangian (sLag) submanifolds are common \cite{1999math......2076B}.  In order for the brane dynamics to be confined to the toric submanifold, we simply need to be in a region of large warping - something that we already require for the flux cascade.  Since branes are attracted to regions of large warping they will be confined to the submanifold and not ``see" the rest of the Calabi-Yau. We are not aware of any example where such toric sLags appear at the bottom of a warped throat but we are also not aware of any argument against their existence.

\begin{figure}
\centering
\includegraphics[width=0.55\textwidth]{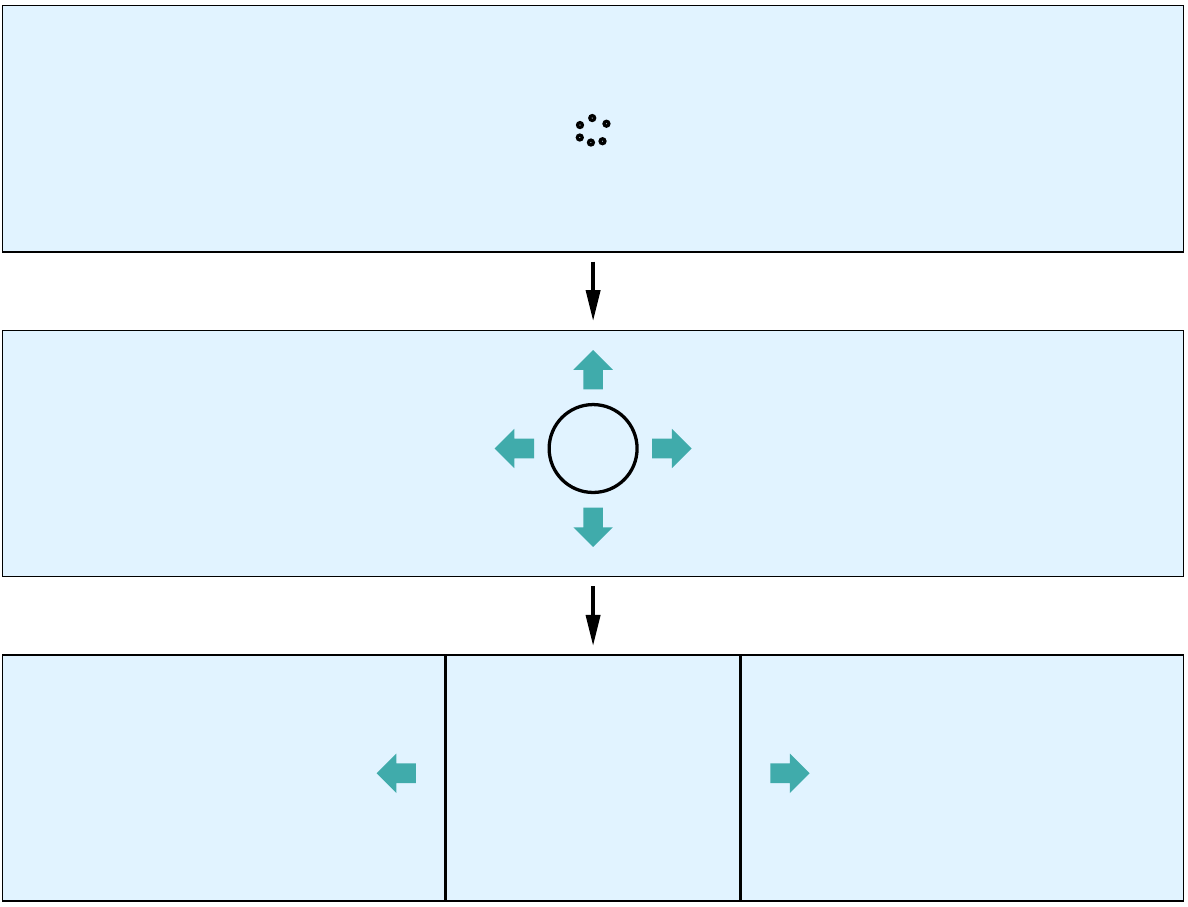}
\caption{\label{toruspic}The polarisation of anti-D3 branes on a long thin torus.  The sides of the  rectangle should be identified to form the torus. A stack of antibranes first forms a spherical D5-brane that grows and collides with itself forming a wrapped pair of D5/anti-D5.}
\end{figure}
For simplicity, we will also consider an anisotropic torus where $L_{1} > L_{2}=L_{3}$.  The reason for this is so that we can consider the case where the anti-D3 branes does not polarize in an isotropic way, but rather forms a brane/antibrane pair that wrap the two-cycle over $T_{2,3}$, but are localized in the $T_1$ direction (see \figref{toruspic}).  This will simplify the dynamics of the cascade because the cascade takes place only in the $T_1$ direction where the pair is co-dimension 1.  As shown in \cite{D'Amico:2012ji}, if the anti-D3 branes polarize into a spherical D5 that is localized in all directions on the three-torus, the cascade will continue in all three directions, discharging the flux faster, and resulting in a different power law for the inflationary potential.  This is also an interesting case to consider, however we will stick to the simplest realization here, where the cascade is only in a single direction.

A calculation identical to the one detailed in \secref{theaction} but using the metric on a torus:
\be
ds^2=h^{-1/2}(dx_\mu dx^\mu + h(dT_2^2+dT_3^2))+h^{1/2}(dT_1^2+dr^2 + r^2d\Omega_2^2)~,
\ee 
leads us to the action for the D5/anti-D5 pair:
\be \label{T3action}
\begin{split}
S= -2 {\mu_3 K \over g_s} \int d^4x a^3(t) h^{-1} & \Bigg[  \sqrt{1-h (\partial_t T_1)^2}\sqrt{\({L_{2} \over 2\pi\ell_s}\)^4 {h \over K^2} +\(\frac{p}{K} - \frac{T_1}{L_1}\)^2} \\
&-\({L_2 \over 2\pi\ell_s}\)^2 {h^{1/2} \over K }+ {p \over K} - {T_1 \over L_1}\Bigg]~.
\end{split}
\ee
There is an overall factor of two to count both the D5 position $T_1$, and its anti-D5 partner at $-T_1$.  Additionally, there is a term that should be included to  account for the interaction of the D5/anti-D5 pair.  This can be taken into account by computing the backreaction of the D5 on the torus geometry, and then placing the anti-D5 into the backreacted geometry at the probe level.  This computation mimics the calculation of the D3/anti-D3 interaction in the KS throat of  \cite{Kachru:2003sx} and is the S-dual of the NS5/anti-NS5 interaction in \cite{Danielsson:2016cit}.

Comparison to the usual backreaction due to the presence of a D brane, $ds^2=h^{-1/2}dx_\parallel^2 + h^{1/2}dx_\perp^2$ implies that the backreaction of the D5 gives $h\to h +\delta h$.  Then the system we need to solve is:
\be 
\begin{split}
ds^2 &= h^{-1/2}H^{-1/2}(dx_\mu dx^\mu + h(dT_2^2+dT_3^2)) +h^{1/2}H^{1/2}(dT_1^2+dr^2 + r^2d\Omega_2^2)~,\\
F_7 &= g_s^{-1}{\rm d}(hH)^{-1} \w \vol_\parallel~, \\
\end{split}
\ee
where the harmonic function, $H=H(r,T_1) \sim (1+ \delta h/h)$, cannot be the usual harmonic for the D5-brane because of the periodicity in the $T_1$ direction.  The periodic harmonic that solves this system is:
\be
H = 1+{\pi (2\pi \ell_s)^2 \over h L_1 r} {\sinh\({2\pi r \over L_{1}}\) \over \cosh\({2\pi r \over L_{1}}\) - \cos\({2\pi T_1 \over L_{1}}\)}~.
\ee
The new action should now be evaluated using the potential resulting from back reaction, $C_6 = (hH)^{-1}$.  Evaluating at $r=0$ we find:
\be 
\mu_5 \int C_6 =  {\mu_3 K\over g_s} \int \dd^4x \,a^3(t)h^{-1} \({L_2 \over 2\pi \ell_s}\)^2{h^{1/2}\over K} \f{ \sin^2\(\pi T_1 \over L_{1}\)}{\big({\pi\over L_1}\big)^2 (2\pi\ell_s)^2+  h \sin^2\(\pi T_1 \over L_{1}\)}~.
\ee
This backreacted value for the potential $C_6$ can be substituted into \eqref{T3action} to include the mutual attraction between the brane and the antibrane

We now pass to a to a canonically normalized field.  As in \secref{cannormsphere} we will expand to second order in velocity and keep next to leading order terms in $p/K$ as long as they do not multiply higher derivative terms.  This leads to a canonically normalized field: $\phi = \sqrt{2\mu_3 p/g_s} T_1 \equiv f T_1/L$, with the action:
\be
S=\int \,d^4x\, a^3(t)\left[ \frac12 \dot{\phi}^2 - {2Kf^2 \over hp L_1^2}\({p\over K}-{\phi\over f}-\({L_2 \over 2\pi \ell_s}\)^2{h^{1/2}\over K}
\f{ \sin^2\({\pi \phi \over f}\)}{\big({\pi\over L_1}\big)^2 (2\pi\ell_s)^2+  h \sin^2\({\pi \phi \over f}\)}\)\right]~.
\ee
We see that there are small periodic perturbations to the linear potential coming from the interaction terms.  These deviations from the linear potential are necessarily small due to both the supergravity approximations where $L_1 \gg \ell_s$, and because the size of the cycle should be fixed by the flux number $h \sim K^2 \gg 1$.  In this case we see that the oscillations are negligible and there is no obstruction to the standard slow roll scenario with a linear potential.  This is the same potential as Axion Monodromy - linear with tunably small periodic perturbations.

\section{Discussion and outlook} \label{discuss}

We have argued that the mechanism of unwinding inflation \cite{D'Amico:2012ji, D'Amico:2012sz} can be embedded in well-known compactifications of type IIB string theory. The essential mechanism relies on the perturbative annihilation of antibranes against the surounding fluxes at the bottom of a warped throat. The inflationary mechanism we point out is based on generic ingredients of flux compactifications and seems rather natural. The inflaton corresponds to the position of five-branes that moves back and forth over a compact cycle discharging a fixed amount of flux in each period of that motion. Therefore the inflaton range is not strickly bounded in the same way as the axion-monodromy models \cite{Silverstein:2008sg, McAllister:2008hb} and large field inflation is possible.

We study this mechanism in two throat geometries, the first is the well known Klebanov-Strassler solution, whereas the second is a more speculative throat containing a $T^3$ at its tip.  The four-dimensional effective potentials that we find in these two cases are:
\bea
V_{S^3}(\phi) &=&  2A_0\e^{4A}\(\frac{\pi p}{K}- {\phi \over  f} + \frac12 \sin\({2\phi \over 
f}\)\)~, \label{potential1} \\
V_{T^3}(\phi) &=& {2Kf^2 \over hp L_1^2}\({p\over K}-{\phi\over f}+ {\cal O}(1/K) \times \text{oscillations}\)~.
\eea
In both geometries we find a potential that is linear plus oscillatory corrections as was previously found in axion monodromy models \cite{Silverstein:2008sg, McAllister:2008hb}. Despite this similarity the underlying mechanisms are not the same.  In particular, the inflaton in this case is not an axion, but rather a D-brane modulus. The geometry is also different: whereas the dynamics in the flux cascade take place on a three-cycle at the bottom of a throat, the axion monodromy scenarios employ two related throat regions (the bifid throat) that have homologous two-cycles at the bottom. However, the relation between these models is not fully understood and is something that we would like to explore in future work.

Using the fully stabilized scenario in the KS throat we are able to achieve a 60 efold inflationary period in which the inflaton has a trans-Planckian field range, but we have yet to find a set of parameters that are also consistent with CMB observations.  Relaxing our control of the geometry and using a speculative toric throat we find no obstacles in finding observationally valid large field inflation. As this is the first string theory embedding of the flux cascade, there remain many open questions which are outside the scope of this work.  We list a few here:
\begin{itemize}
\item As mentioned in the text, we find that inflation ends when the positive energy from the anti-D3 branes is no longer large enough to compensate for the negative vacuum energy for the K\"ahler modulus $\sig$.  If the cascade continues past this point, inflation will end in AdS and we would need to posit some unknown phase transition or uplift mechanism in order to restore de Sitter space.  However, there is some reason to hope that dissipative effects will stop the cascade before all the anti-D3 charge is gone. These dissipative corrections should come from open string production and closed string bremsstrahlung \cite{McAllister:2004gd, D'Amico:2014psa, Bachlechner:2013fja}.  While there are no current estimates for these effects for spherical branes or that apply in the presence to RR fields, we expect that these effects become important where the acceleration becomes large.  Indeed, we find a spike in the acceleration directly before the total vacuum energy becomes negative.  Additionally, one might worry that open string production at the poles of the sphere is large enough to immediately stop the cascade.  However, the fact that the cascade takes place at non-relativistic velocities means that open string masses should be changing adiabatically, suppressing string production.  Furthermore, because the D5 brane is carrying anti-D3 charge, it cannot simply annihilate at the poles as one might expect for a spherical brane. 
\item By going beyond the probe approximation for the five-brane, one can potentially reduce the amplitude of oscillations in \eqref{potential1}  . Corrections to the probe potential should come in powers of $R_{\rm D3}/R_{\rm cycle}$.  As argued in \cite{DeWolfe:2004qx}, the probe potential breaks down near the poles of the three-sphere\footnote{The potential at small angles should be computed as in Polchinski-Strassler \cite{Polchinski:2000uf} via the world-volume theory on the system of non-Abelian anti-D3 branes perturbed by fluxes.
} where the oscillations are most prominent.  The arguments in  \cite{Bena:2014jaa,Bena:2016fqp,Danielsson:2016cit} suggest that backreaction suppresses the oscillations.  The reasoning is that the predicted corrections are such that the tendency to create meta-stable states is lost. 
\item It is possible that the anti-D3 branes polarize into multiple D5 branes instead of just one\footnote{We thank Iosif Bena for bringing this point to our attention}.  These multiple D5 channels are energetically unfavorable, and so we have neglected them here.  However the kinetic energy of the D5 during the cascade could cause these channels to be populated.  This would result in an altered inflationary potential because more units of flux, and therefore more antibrane charges, would be discharged in each step of the cascade. 

\item There remain some aspects of moduli stabilization that are not well under control. First, the relation between the K\"ahler modulus $\sigma$ and the warped volume of the Calabi-Yau is non-trivial, however the two are not independent.  In the absence of warping one finds $\sigma \sim {\cal V}^{2/3}$.  Using this as an estimate we see that the warped volume in table \ref{params} exceeds this value by many orders of magnitude.  Second, the flux numbers mentioned in table \ref{params} are in conflict with the known Euler numbers for elliptically fibered fourfolds.  Although there may exist a Calabi-Yau with an Euler number large enough to accommodate this amount of flux, it would be preferable to work with an Euler number that is known to exist. Alternatively, we could arrange the parameters such that after inflation $KM$ is significantly reduced. This would require the majority of the fluxes to be discharged by the cascade. This change in fluxes would backreact in an important way on the throat geometry and the flux superpotential. Both of these issues require further study and probably necessitate going beyond the simplest single-K\"ahler modulus stabilization mechanism of KKLT.

\end{itemize}

\acknowledgments
We are very grateful to Iosif Bena, Liam McAllister, Bert Vercnocke and Alexander Westphal for detailed feedback on the manuscript. We are also happy to acknowledge useful discussions with Thomas Bachlechner, Matthew Kleban, Dario Martelli, Nikolay Bobev, Guido D'Amico, Gary Shiu and Eva Silverstein.
The work of FFG was supported by the John Templeton Foundation Grant 48222. The work of FFG and TVR is supported by the FWO odysseus grant G.0.E52.14N.  The work of MS is supported by the European Union's Horizon 2020 research and innovation programme under the Marie Sk{\l}odowska-Curie grant agreement No 656491.  We acknowledge support from the European Science Foundation HoloGrav Network, the Belgian Federal Science Policy Office through the Inter-University Attraction Pole P7/37, and the COST Action MP1210 `The String Theory Universe'.

\bibliography{refs}

\end{document}